\documentstyle[epsf,pre,aps,twocolumn]{revtex}
\newcommand{\ul}[1]{\underline{#1}}
\newcommand{\ld}{{\tiny\ldots}}
\begin{document}

\draft
\title{Comment on ``Universal formulas for percolation
  thresholds. II. Extension to anisotropic and aperiodic lattices''}

\author{F. Babalievski\thanks{Permanent address:
Institute of General and Inorganic Chemistry,
 Bulgarian Academy of Sciences, 1113 Sofia, Bulgaria}}

\address{ Institute for Computer Applications 1 (ICA1)\\
    University of Stuttgart, 70569 Stuttgart, Germany}
\date{November 17, 1997}
\maketitle
\begin{abstract}
Recently S. Galam and A. Mauger $[$Phys. Rev. E {\bf 56}, 322 (1997)$]$
proposed an
approximant which relates the bond and the site percolation threshold
for a particular lattice. Their   formula is based on a fit to exact and
simulation
results obtained earlier for different periodic and aperiodic lattices. 
However, the numerical result for an aperiodic dodecagonal
lattice does not agree well with the proposed formula.
I present here new and more precise data for this and other
aperiodic lattices. The previously published value for the
dodecagonal lattice is confirmed. The reason for the deviation 
from the Galam and Mauger approximant is discussed.
  
\end{abstract}


In a recent article \cite{GM97b} S. Galam and A. Mauger proposed a
formula which relates  the site and bond  percolation
thresholds for a given lattice.  This formula is a
modification
of previous one \cite{GM96} designed to predict the percolation
thresholds for a variety of periodic lattices. 
In the new Galam and Mauger approximant (NGMA) the (mean) coordination
number of a lattice ($z$) is replaced by  effective one 
($z_{\text{\it eff}}$):

$
p_c = p_0\, [(d-1)(z_{\text{\it eff}}-1]^{-a}d^{\,b}
$

\noindent
where $p_0$, $a$ and $b$ are free parameters and the (countable) set of values
for
them divide lattices in separate classes. The formula provides a
connection  between the
site and the bond percolation threshold in a particular class. So, if
(say) 
the bond percolation
threshold is known for a lattice (and one guesses somehow to which class
belongs this lattice) then one can determine $z_{\text{\it eff}}$. After replacing
 $p_0$, $a$ and $b$ with their values for site percolation 
(within the same class) one could obtain a value for the site
percolation threshold. In the same manner one could start from a site
percolation threshold and to get the bond threshold.

The  NGMA was tested  on simulation and exact values for 17 lattices
and among them six 
lattices based on aperiodic tilings. The values for two variants of
the octagonal and dodecagonal aperiodic lattices  were
extracted from my computer simulation results \cite{PhysA95}. 
  It appeared that some of these values do not fit well to the NGMA.
The authors of \cite{GM97b} supposed that the value for bond
percolation on the ferro variant of the dodecagonal lattice should 
be corrected. Instead of my result $p_c = 0.495$ they suggest that the
correct value should be $0.475$ -- so it would fit much better to the
theoretical curve (the NGMA). Since the bond percolation
results given in my paper were stated to be preliminary, it is
reasonable 
to be suspicious about them.

Here I present new and more precise numerical data for these thresholds
as well as for the bond percolation thresholds for the other aperiodic
lattices used for testing the NGMA  -- see the left part of 
Table \ref{Tablepc}.

These values were obtained by computer simulation analogous to that
in \cite{PhysA95}. The lattice sizes were much larger here: up to size 
$500 \times 500$ bond lengths (the longer bond)  and the computational
efforts were
 between 100 and 1000 times larger for the different lattices. For
 this work  I
used other  pseudo random number generators (PRNG).
For the most part of simulation the {\tt drand48()} PRNG from the
standard GnuC distribution was used. I used also the {\tt ran2} generator
from \cite{numrecC} and the recently discussed \cite{Ziff97r} 
four tap shift-register generator $R(471,1586,6988,9689)$. 
The results coincided within less than $1/10$ of a percent
in the mean of the cumulative Gaussian distribution which
was used for a fitting function. The fits just with a polynomial of
3-th degree gave almost the same results for the places (the values of
$p$) where the
spanning probability curve reaches a value of 1/2. One could pay
attention also, that the value for the non-modified Penrose lattice
conforms the earlier results.  

So I could claim (within the rigor of a Monte Carlo simulation) that
the  bond percolation threshold for the   ``ferro'' modification of the
dodecagonal aperiodic lattice is indeed $0.495$ within error bars of
$\pm 0.001$. In fact a more precise value could be extracted from the
data on Fig.1 $p_c=0.4950\pm 0.0005$. The error bars would be even
smaller if one disregard the shift in the spanning probability
curve for $L=300$.

Now one could draw the conclusion that  deviations of this
percolation
threshold from the NGMA do exist and one has either to accept that
the approximant is good within a larger interval of deviations: $|p_c
-p_c^e|=\Delta <0.014$ or to search a way to improve the approximant
itself.  

The reason for the discrepancy could be suggested in the line of
thinking  which  S. van der Marck has presented recently
 \cite{SMarck97c} for site percolation. He paid attention that for
 certain  lattices there are 
sites positioned on specific positions which do not contribute to  
percolation at all --- e.g. in 2D these are the ``star sites'' inside
a triangle. 

 Indeed for bond percolation  such entirely irrelevant bonds are 
 impossible. However,
 one can show that some bonds contribute much less to the short range
 connectivity than the others. This is the case of the questioned
aperiodic dodecagonal lattice with ``ferro'' bonds. This lattice is
a modification of the original dodecagonal lattice constructed by
Socolar, which corresponds to an aperiodic tiling of plane with
hexagon, square and a rhombus. The ``ferro'' modifications consists
in adding new bonds along the short diagonals of the rhombuses. Now
one can show that these new bonds contribute less to the short range
percolative connectivity.

An estimation for this contribution could be done by stating a
percolation
problem for ``one rhombus system''.  We have five bonds arranged in a 
rhombus where the fifth bond is along the  short diagonal. The
question
is what is the probability for a bond to be a ``cutting'' one for
percolation between sites on the acute angles (i.e. what is the
probability that the extraction of this bond destroys the connection between
the two sites). One can easily see that this probability is 
$2 p^3(1-p)^2$ (the two zig-zag paths) for the ferro bond, which gives 
$\approx 1/16$ for $p \approx 1/2$ (close to the questioned threshold
value). The same value  (near to $p=1/2$) is $\approx  3/16$ for the
other bonds $[4 p^3 (1-p)^2 + p^2(1-p)^3 +p^4(1-p)]$. 

Of course, things change when one considers percolation between the 
sites connected directly with the ``ferro'' bond. 
The cutting probability (again near $p=1/2$) for this bond is already 
$\approx 9/32$ $[p(1-p^2)^2]$
 and the cutting probability of each one of the other bonds
is  $\approx 3/32$  $[p^2(1-p)^2 (1+p)]$.

It appears that the  chance to be a cutting bond is  the
same  for the two type of bonds if  both pairs of
opposite vertices of the rhombus are considered . But  one has to
mention (e.g. on Fig. 1{\bf g} in \cite{GM97b}): 
there are only one or two outgoing bonds from the obtuse angle vertices,
while the acute vortices have 2 to 4 outgoing bonds. 

So if one considers the ``spreading of  connectivity'' one more step beyond
 the two diagonals of the rhombus, the contribution of the ''ferro''
bond is  less significant (by a factor of two approximately).

In contrast  to the above, the same analysis
for  central-force rigidity percolation\cite{Thorp95} on the same
lattices\cite{ALFB97} 
shows that the
``ferro'' bond is always cutting bond within the questioned rhombus. 
There is a simple approximant for the rigidity thresholds
-- so called Maxwell approximant\cite{Thorp95,ALFB97}. 
It is just twice the other mean-field like approximant:
the Scher and Zallen prediction for (connectivity) bond percolation
thresholds in two dimensions ($p_c \approx 2/z$, where $z$ is the mean
coordination number).
One can see from the Table \ref{Tablepc} that the estimates for
rigidity  percolation models,
where each bond has (almost) equal significance for spreading the
rigidity, agree with the Maxwell
approximant extremely well.  

I thank H.J. Herrmann and J.-P. Hovi for critical reading the manuscript.
This research was supported by the 
German Academic Exchange Foundation (DAAD).

 \begin{figure}
\centering
\epsffile{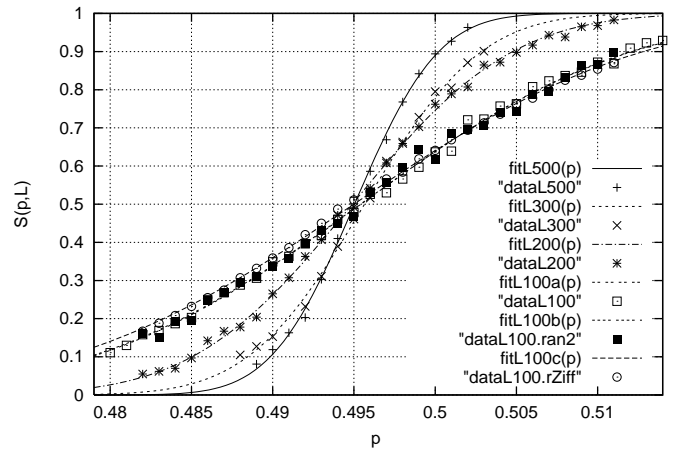}
 \caption{The frequency of occurrence of spanning cluster $S(p,L)$ for 
dodecagonal lattices with   different size($L$). Each data point is
averaged over 1000 realizations.}
\label{dodecspan}
\end{figure}

\begin{table}

\caption{New numerical results for the bond percolation thresholds of
questioned aperiodic lattices. The values in right are the rigidity
(bond) percolation thresholds ($ p_c^r $) taken from $[8]$.
Some of the lattices are always floppy (a.f.) in
view of the rigidity percolation model. One can see that the lattices
problematic for the NGMA have $ p_c $ larger than the half  of the respective
rigidity threshold. The error estimates are put in brackets and concern
the underlined digit.}
\label{Tablepc}
\begin{tabular}{||l|l|c|c|c||}
              & $\; \; p_c^{conn}$ & $2/z$  & $p_c^{rig}$&$4/z$ \\
\hline
 Penrose      & $0$.$476\ul{7}(5)$ &$1/2$   & a.f.         & $1$ \\
\hline
Penrose (f)   &$0$.$42\ul{9}(1)  $ &$0.420$\ld& $0.83\ul{6}(2)$&$0.840$\ld \\
\hline
Octagonal     &$0$.$47\ul{8}(3)  $ &$1/2$   & a.f.         & $1$ \\
\hline   
Octagonal(f)  &$0$.$40\ul{2}(5)  $ &$0.387$\ld& $0.76\ul{9}(2)$ & $0.774$\ld\\
\hline
Dodecagonal   & $0$.$53\ul{8}(1) $ &$0.551$\ld& a.f.      &$1.102$\ld \ \\
\hline
Dodecagonal(f)& $0$.$495\ul{0}(5)$  &$0.468$\ld& $0.93\ul{8}(1)$& $0.937$\ld\\
 \end{tabular}
 \end{table}
 
\end{document}